\documentclass[aps,pre,showpacs,reprint,groupedaddress]{revtex4-1}

\pdfoutput=1

\usepackage{amsmath}
\usepackage{lmodern}
\usepackage{graphicx}

\usepackage[colorlinks,citecolor=blue]{hyperref}

\usepackage{amssymb} 

\begin{document}

\title{Symmetry breaking in two interacting populations of quadratic integrate-and-fire neurons}
\author{Irmantas Ratas and Kestutis Pyragas}

\affiliation{Center for Physical Sciences and Technology, LT-10257 Vilnius, Lithuania}

\begin{abstract}
We analyze the dynamics of two coupled identical populations of quadratic integrate-and-fire neurons, which  represent the canonical model for class I neurons near the spiking threshold. The populations are  heterogeneous; they include both inherently spiking and excitable neurons. The coupling within and between the populations is global via synapses that take into account
the finite width of synaptic pulses. Using a recently developed reduction method based on the Lorentzian ansatz, we derive a closed system of equations for the neuron's firing rates and the mean membrane potentials in both populations. The reduced equations are exact in the infinite-size limit. The bifurcation analysis of the equations reveals a rich variety of non-symmetric patterns, including a splay state, antiphase periodic oscillations, chimera-like states, also chaotic oscillations as well as bistabilities between various states. The validity of the reduced equations is confirmed by direct numerical simulations of the finite-size networks.

\end{abstract}


\maketitle

\section{\label{sec1}Introduction}

Studies of collective motion in networks of nonlocally or globally coupled oscillators or excitable elements are the focus of current research in diverse fields from physics to neuroscience. Starting from pioneering works of Winfree~\cite{winfree67} and Kuramoto~\cite{kuramoto75} various striking effects such as synchronization, collective chaos, and chimera states have been detected~\cite{Pikovsky2015}.

The existence of chimera states was first reported by Kuramoto and Battogtokh~\cite{kuramoto2002}. They considered a ring of identical nonlocally coupled oscillators and showed that they could spontaneously split into synchronized and desynchronized subpopulations. Though the oscillators were coupled symmetrically and the system possessed a translational symmetry, for the chimera states this symmetry was broken. By now chimera states were detected and analyzed in diverse systems with different types of topology, various types of oscillators and  different types of couplings. For a recent review of the subject we refer to Ref.~\cite{panaggio2015}. 

A major breakthrough in an analytical treatment of chimera states was achieved by Abrams et al~\cite{abrams08}. They considered the simplest setup that supports chimera states: a
pair of oscillator populations in which each oscillator is coupled equally to all the others in its group, and with different strength to those in the other group. Such a system is symmetric with respect to exchange of populations and here chimera states represent the symmetry broken solutions with the coherent and incoherent behavior of oscillators in different populations. The authors solved the problem in the infinite-size (thermodynamic) limit by applying Ott and Antonsen ansatz~\cite{ott08}. They derived a simple system of two ordinary differential equations that characterize the macroscopic dynamics of the network and obtained the exact results about the stability, dynamics, and bifurcations of chimera states. Inspired by the work~\cite{abrams08}, different authors employed the setup with two globally coupled populations to analyze chimera states in a large variety of models~\cite{laing2009,pazo14,Gonzalez2014,Pinaggio2016,olmi2010,bolotov2016,shepelev2017}.

Chimera states are of particular interest in neural models~\cite{olmi2010,bolotov2016,shepelev2017,sakaguchi2006,omelchenko2013,hizanidis2014,omelchenko2015,
bidesh2016,bera2016b,glaze2016,semenova2016,isele2017}.  Many creatures like birds, reptiles and sea mammals sleep with only half their brain at a time~\cite{Rattenborg2008}. In such a unihemispheric sleep the awake side of the brain shows desynchronized electrical activity, whereas the sleeping side is highly synchronized. The authors of Ref.~\cite{abrams08} suggested that chimera states in two coupled neural populations might serve as a model of unihemispheric sleep. An asymmetric brain activity has been also observed in human sleep apnea patients~\cite{Rial2013}. Thus  the study of neural chimera states may have clinical relevance as well.

In this paper, we analyze two globally coupled populations of quadratic integrate-and-fire (QIF) neurons. The isolated  QIF neuron is the canonical model for the class I neurons near the spiking threshold. The spiking instability in such neurons appears through a saddle-node bifurcation on an invariant curve (SNIC), in which a pair of fixed points on a closed curve coalesce to disappear, converting the curve to a periodic orbit. The peculiarities of our model are as follows. Unlike to typical models, which consider chimera states in systems of identical oscillators, here we analyze two populations of heterogeneous neurons. Each population contains both excitable and spiking neurons. The interactions between neurons are provided either by synaptic coupling or mean field potential. Our model admits an analytical treatment via a recently developed Lorentzian ansatz (LA) method \cite{montbrio15,ratas2016}. In the thermodynamic limit, we derive a simple system of ordinary differential equations, which describe the macroscopic dynamics of the firing rates and  mean membrane potentials in both neural populations. This macroscopic model enables us to perform a thorough bifurcation analysis of the system. As a result, we detect two types of chimera-like states. In one of them, the majority of neurons in one population are quenched, while in another population they spike synchronously. In the second type, the majority of neurons produce spikes in both populations, but with a different synchronization level.  

The paper is organized as follows. The microscopic model of two synaptically coupled populations of QIF neurons and derivation of macroscopic model equations in the thermodynamic limit is described in Sec.~\ref{sec2}. Section~\ref{sec3} is devoted to the stability analyzes of symmetric solutions of the macroscopic model. The analysis of non-symmetric solutions and their bifurcations is presented in Sec.~\ref{sec4}. In Sec.~\ref{sec5} we consider the case when the coupling between populations is defined by a mean field potential rather than the synaptic interaction. The paper is concluded by discussion in Sec.~\ref{sec6}.

\section{\label{sec2}The Model}

Our consideration of two neural populations is based on the heterogeneous model of all-to-all synaptically coupled quadratic integrate-and-fire neurons, which are the canonical representatives for a class I neurons near the spiking threshold. One such population has
been thoroughly studied in Ref.~\cite{ratas2016}. The membrane potential of each neuron $V_j$, (here ${1\leq j \leq N}$, $N$ is the size of the population) is described by the following equation \cite{ermentrout10}:
\begin{equation}
\dot{V}_{j} =  V_{j}^{2}+\eta_{j}+I^{syn}_{j}.\label{model}
\end{equation}
Here the  constants $\eta_j$ specify the behavior of  individual neurons. For $\eta_j<0$ the neuron is in an excitable regime and for $\eta_j>0$ it is in the spiking regime. We assume that the values of the parameters $\eta_j$ are distributed according to some defined density function $g(\eta)$ and that system (\ref{model}) contains both excitable ($\eta_j<0$) and spiking neurons ($\eta_j>0$). Whenever the membrane potential $V_j$ reaches the peak value $V_{peak}$ its voltage is reset to the value  $V_{reset}$. In order to treat the system \eqref{model} analytically, the peak and reset voltages are set to infinity ${V_{peak}=-V_{reset} \to \infty}$. 

The term $I^{syn}_{j}$ stands for the synaptic current. We assume all-to-all homogeneous neural coupling and write this term in the form:
\begin{equation}
I_j^{syn} = - K (V_j-V_s)\frac{1}{N}\sum_{l=1}^{N} s_l.  \label{syn_curr}
\end{equation}
Here $K$ is the maximal conductance of postsynaptic receptors and $V_s$ is the reversal potential of synapse. In the case of fast synaptic processes, the fraction of open ion channels in the neuron membrane is described by sigmoid function, $s_l = \{1+\exp[-\sigma(V_l-V_{th})] \}^{-1}$, with steepness parameter $\sigma$ and threshold potential $V_{th}$. This oft-used coupling
form is called fast threshold modulation \cite{somers1993}. We consider the limiting version $\sigma \to 0$, when the sigmoid function transforms into the Heaviside step function, $s_l=H(V_l-V_{th})$. Moreover, to reduce the number of parameters, we consider the limit $V_s \to \infty$ and $K \to 0$ with the product  $KV_s$ remaining finite. Then by defining the new parameter $J =KV_s/V_{th}$, we obtain the simplified expression for the synaptic current that does not depend on the index $j$, $I_j^{syn}=I^{syn}$,
\begin{equation}
	I^{syn}=J\frac{V_{th}}{N}\sum_{l=1}^{N}H(V_l-V_{th}). \label{syn_curr1}
\end{equation}
This expression is a good approximation for small excitatory synapses on a large compartment \cite{Roth2009}. In that case, the depolarization of the membrane is small and the difference $V_j-V_s$ is little changed during the excitatory postsynaptic potential. Note that the latter approximation is not necessary for the analytical treatment of the model. The reduced system of macroscopic equations for the neuron's firing rate and the mean membrane potential can be derived in the thermodynamic limit without recourse to this approximation \cite{ratas2016}. However, below we will use this approximation, since it simplifies the bifurcation analysis of the macroscopic equations. Moreover, as shown in Ref.~\cite{ratas2016}, the bifurcation diagrams obtained with the synaptic currents \eqref{syn_curr} and  \eqref{syn_curr1} are qualitatively similar.  In both cases the model demonstrates a large variety of dynamical regimes, including a single steady state solution, bistability between two different steady states, macroscopic self-oscillations as well as bistability between the steady state and self-oscillations. 

In this paper we consider two interacting populations of neurons of the described type
\begin{equation}
	\dot{V}_{j,k}=V^2_{j,k}+\eta_{j}+\mathcal{I}_{k}. \label{modelC}
\end{equation}
We assume that the populations are identical. The index $j=1,\ldots,N$ labels the neurons inside of each population, while the index $k=0,1$ marks the populations. The term $\mathcal{I}_{k}$ describes both the interaction between the neurons inside of each population and between the populations. In this paper we mainly focus on the analysis of the situation  when the internal and external interactions between neurons are modeled by the synaptic current in the form \eqref{syn_curr1}. The expression for term $\mathcal{I}_{k}$ in this case is 
\begin{equation}
\mathcal{I}_{k} = \left( J_{in}  S_{k}+J_{ex} S_{1-k} \right) V_{th},\label{Isyn}
\end{equation}
where 
\begin{equation}
S_k = \frac{1}{N}\sum_{l=1}^{N} H(V_{l,k}-V_{th}) \label{Sdef}
\end{equation}
and the parameters $J_{in}$ and $J_{ex}$ define the coupling strengths within and between the populations, respectively. In Sec.~\ref{sec5}, we briefly describe another situation, when the interaction between the populations is provided by a mean field coupling rather than the synaptic coupling. 

\subsection{Thermodynamic limit}

In the thermodynamic limit $N \to \infty$, the system  \eqref{modelC} of two interacting neural populations can be reduced to a system of only four ordinary differential equations, which defines the dynamics of firing rates and mean membrane potentials of individual populations. Such macroscopic equations can be derived by a recently developed reduction method based on the Lorentzian ansatz. The idea of this method has been first proposed for a single network of QIF neurons interacting via instantaneous  pulses \cite{montbrio15} and then extended to a more realistic model of synaptic interaction that takes into account the finite width of synaptic pulses \cite{ratas2016}. The technique of the derivation of the macroscopic equations for the  system  \eqref{modelC} is similar to that described in Refs. \cite{montbrio15} and \cite{ratas2016}, and thus we present this derivation in abbreviated form.

In the infinite-$N$ limit,  the macroscopic state of each population in the system~\eqref{modelC} can be described by  continuous density functions $\rho_k(V|\eta,t)$, $k=0,1$.  The product $\rho_k(V|\eta,t) \text{d} V$ defines the fraction of neurons in the $k$th population with the membrane potential between $V$ and $V+\text{d} V$ and parameter $\eta$ at time~$t$. These density functions satisfy continuity equations
\begin{equation}
\frac{\partial}{\partial t}\rho_k = -\frac{\partial}{\partial V} \left[ \rho_k \left \{ V^2+\eta +\mathcal{I}_{k} \right \} \right], \label{eq_cont}
\end{equation}
where $\mathcal{I}_{k}$ is defined in Eq.~\eqref{Isyn}. In the continuous limit, the sum $S_k$ defined in Eq.~\eqref{Sdef} becomes a double integral
\begin{equation}
S_k=\int\limits_{-\infty}^{+\infty} g(\eta) \int\limits_{-\infty}^{+\infty}  \rho_k(V|\eta,t) H(V-V_{th}) \text{d}V \text{d}\eta. \label{St} 
\end{equation}
The main assumption of the LA ansatz is that the solutions of Eqs.~\eqref{eq_cont}, for any initial conditions, converge to a Lorentzian-shaped function (see Ref.~\cite{montbrio15} for the relation between the LA ansatz and Ott-Antonsen ansatz~\cite{ott08})
\begin{equation}
\rho_k(V|\eta,t)=\frac{1}{\pi} \frac{x_k(\eta,t)}{[V-y_k(\eta,t)]^2+x_k(\eta,t)^2},\label{eq_LA}
\end{equation}
where time-dependent parameters $x_k(\eta,t)$ and $y_k(\eta,t)$ define the half-width  and the  centre of the distribution. The parameters $x_k(\eta,t)$ and $y_k(\eta,t)$  characterize all relevant dynamics of the system in a reduced subspace. They are related to the total firing rate $r_k(t)$ and the mean membrane potential $v_k(t)$ via integrals 
\begin{subequations}
\label{eq_rv}
\begin{eqnarray}
r_k(t) & = & \frac{1}{\pi} \int_{-\infty}^{+\infty} x_k(\eta,t) g(\eta) d\eta,\label{eq_rint}\\
v_k(t) & = & \int_{-\infty}^{+\infty} y_k(\eta,t) g(\eta) \text{d}\eta.\label{eq_vintgrl}
\end{eqnarray}
\end{subequations}

Substituting the LA \eqref{eq_LA} into the continuity Eqs.~\eqref{eq_cont}, one can derive a system of differential equations for $x_k(\eta,t)$ and $y_k(\eta,t)$
\begin{subequations}
\label{eq_xy1}
\begin{eqnarray}
\dot{x}_k(\eta,t) & = & 2x_k(\eta,t)y_k(\eta,t),\label{eq_xy1a}\\
\dot{y}_k(\eta,t) & = & \eta-x_k^2(\eta,t)+y_k^2(\eta,t) + \mathcal{I}_{k},  \label{eq_xy1b}
\end{eqnarray}
\end{subequations}
which for the complex variable $w_k(\eta,t)\equiv x_k(\eta,t) +i y_k(\eta,t)$ can be written as
\begin{equation}
\dot{w_k}(\eta,t)= i \left[ \eta -w_k^2(\eta,t) + \mathcal{I}_{k} \right].\label{eq_w}
\end{equation}
A simplification can be gained by choosing the density distribution of the $\eta$ parameter in the Lorentzian function form
\begin{equation}
g(\eta)=\frac{1}{\pi} \frac{\Delta}{(\eta-\bar{\eta})^2+\Delta^2}\label{eq_lorz}
\end{equation}
with the width $\Delta$ and the center at $\bar{\eta}$. In this case the integrals \eqref{St} and \eqref{eq_rv} can be solved by extending $\eta$ to the complex plane and computing a contour integral over an infinitely large semicircle in the lower half-plane \cite{montbrio15}. The values of these integrals are defined by the pole ${\eta=\bar{\eta}-i \Delta}$ of $g(\eta)$ function. This enables us to relate the complex variable $w_k$ with the firing rate and the mean membrane potential
\begin{equation}
\pi r_k(t)+ i v_k(t)= w_k(\bar{\eta}-i \Delta,t) \label{eq_rvw}
\end{equation}
as well as obtain the explicit expression for the synaptic function
\begin{equation}
S_k(t)=\frac{1}{\pi} \left[\frac{\pi}{2}- \arctan \left(  \frac{V_{th}-v_k(t)}{ \pi r_k(t)}  \right)\right]. \label{St2}
\end{equation}
Taking into account Eqs.~\eqref{eq_w} and~\eqref{eq_rvw}, the  firing rates and the mean membrane potentials satisfy the system of four differential equations
\begin{subequations}
\label{eq_rv2}
\begin{eqnarray}
\dot{r}_k & = & \Delta/\pi+ 2r_k v_k,\label{eq_r1}\\
\dot{v}_k & = & \bar{\eta} +v_k^2-\pi^2 r_k^2+\mathcal{I}_k, \label{eq_v1}
\end{eqnarray}
\end{subequations}
where $k=0,1$. These equations together with Eqs.~\eqref{Isyn} and \eqref{St2} form the closed
macroscopic model for the network consisting of two synaptically coupled populations of QIF neurons.

\section{\label{sec3}Symmetric solutions and their stability}

The macroscopic Eqs.~\eqref{eq_rv2}  possess permutational symmetry: they are invariant under the change of variables $(r_0,v_0,r_1,v_1) \to (r_1,v_1,r_0,v_0)$. This symmetry admits the existence of the symmetric solutions $(r_1,v_1) = (r_0,v_0)$. To analyze the stability of such solutions, it is convenient to introduce new variables
\begin{subequations}
\label{newCoord}
\begin{eqnarray}
R & = & (r_0-r_1)/2,\\
P & = & (v_0-v_1)/2,\\
Q & = & (r_0+r_1)/2,\\
M & = (& v_0+v_1)/2,
\end{eqnarray}
\end{subequations}
where $(R,P)$ and $(Q,M)$ are the transverse and longitudinal coordinates, respectively. In the new coordinates $(R,P,Q,M)$, the trajectories of the symmetric solutions are placed in the invariant subspace $(0,0,Q,M)$, where the variables $Q$ and $M$ satisfy differential equations
\begin{subequations}
\label{eqQM}
\begin{eqnarray}
\dot{Q} & =&\Delta/\pi+2QM,\\
\dot{M} & = & \bar{\eta}+M^{2}-\pi^{2}Q^{2}+(J_{in}+J_{ex})V_{th} \nonumber \\
&\times & \frac{1}{\pi}\left[\frac{\pi}{2}-\arctan\left(\frac{V_{th}-M}{\pi Q}\right)\right].
\end{eqnarray}
\end{subequations}
These equations are identical to the equations that describe the dynamics of a single population of QIF neurons with a modified coupling strength $J=J_{in}+J_{ex}$. The solutions of this system have previously been analyzed in Ref.~\cite{ratas2016}. It has been shown that the system~\eqref{eqQM} has two types of asymptotically stable solutions: fixed points (steady states) and limit cycles. For some values of the parameters, these solutions can coexist in different combinations giving rise to the bistability. The stable solutions of the system ~\eqref{eqQM} constitute longitudinally stable solutions of the system~\eqref{eq_rv2} in the invariant subspace $(0,0,Q,M)$, while their transverse stability requires a special analysis.

The transverse stability of the symmetric solutions is defined by the variational equations of the $(R,P)$ variables
\begin{equation}
\left(\begin{array}{c}
\delta\dot{R}\\
\delta\dot{P}
\end{array}\right)=\boldsymbol{A}\left(\begin{array}{c}
\delta R\\
\delta P
\end{array}\right)\label{eq:RPdev}
\end{equation}
with the matrix
\begin{equation}
\boldsymbol{A} = 2\left(\begin{array}{cc}
M & Q\\
-\pi^{2}Q+S_{M}(M-V_{th}) & M-S_{M}Q
\end{array}\right) \label{eqAmatrix}
\end{equation}
and the parameter
\begin{equation}
S_M = \frac{V_{th}(J_{ex}-J_{in})}{2\left[\pi^2 Q^2+(M-V_{th})^2\right]}  \label{eqSM}.
\end{equation}
The variables $Q$ and $M$ in Eqs.~\eqref{eqAmatrix} and \eqref{eqSM} satisfy Eqs.~\eqref{eqQM}.  Equation \eqref{eq:RPdev} governs the dynamics of transverse deviations $(\delta R, \delta P)$ from the invariant subspace. When these deviations decay in time than the corresponding symmetric solution defined by Eq.~\eqref{eqQM} is transverse stable, otherwise it is unstable. 

The analysis of transverse stability of  fixed points and limit cycles is different. For the fixed points, the matrix $\boldsymbol{A}$ is constant and its eigenvalues $\lambda_{1,2}$ define the stability. If $\mathrm{Re} \lambda_{1,2}<0$, the symmetric fixed point solution of the system~\eqref{eq_rv2} is transverse stable. For the limit cycles, the matrix $\boldsymbol{A}$ depends on time periodically, $\boldsymbol{A}(t)=\boldsymbol{A}(t+T)$, where $T$ is the period of the limit cycle. In this case we recourse to the Floquet theory. We solve differential equations for the fundamental matrix $\boldsymbol{\Phi}(t)$: $\dot{\boldsymbol{\Phi}}(t)=\boldsymbol{A}(t) \boldsymbol{\Phi}(t)$ with the initial condition equal to the identity matrix, $\boldsymbol{\Phi}(0) =\mathbb{I}$. Then we compute the eigenvalues $\mu_{1,2}$ of the monodromy matrix $\boldsymbol{\Phi}(T)$. If $|\mu_{1,2}|<1$, the symmetric limit cycle solution of the system~\eqref{eq_rv2} is transverse stable. Note that the transverse stability of the symmetric solutions depends on the difference $J_{ex}-J_{in}$, while the solutions themselves are defined by the sum $J_{ex}+J_{in}$.

\begin{figure}
\centering\includegraphics{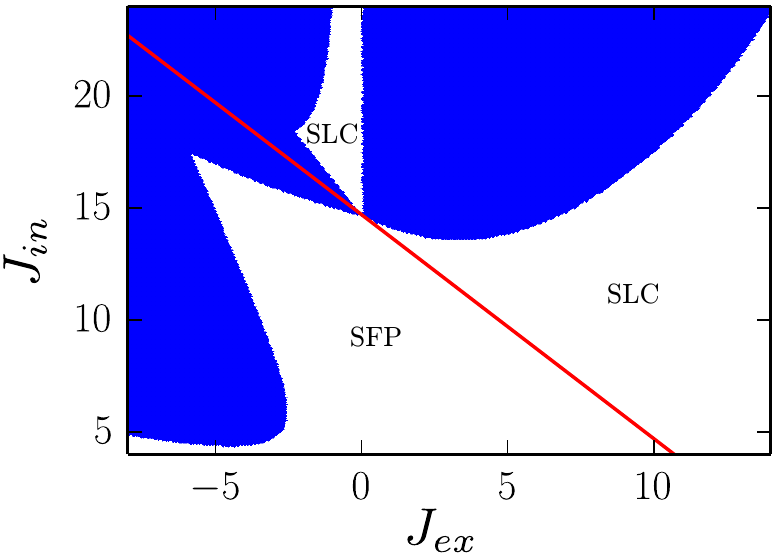}

\caption{\label{fig_JexJinB} Stability diagram of the symmetric solutions of Eqs.~\eqref{eq_rv2} in the plain of parameters $(J_{ex},J_{in})$ for $\bar{\eta} = 0$, $V_{th} = 50$ and $\Delta=1$. The red diagonal indicates the Hopf bifurcation of the system~\eqref{eqQM}. Below this diagonal there are symmetric fixed points (SFP) and above the diagonal --- symmetric limit cycles (SLC). In the white regions, these symmetric  solutions  are transverse stable, while in the blue regions they are unstable and here there are no symmetric attractors.} 
\end{figure}

In Fig.~\ref{fig_JexJinB}, we show the results of the above stability analysis in the plain of parameters $(J_{ex},J_{in})$. The values of the parameters $\bar{\eta} = 0$, $V_{th} = 50$ and $\Delta=1$ are chosen so that the system ~\eqref{eqQM} is monostable for any values of $J_{ex}+J_{in}$. The red diagonal line  $J_{ex}+J_{in} = J_H$ in Fig.~\ref{fig_JexJinB} indicates the Hopf bifurcation of the system ~\eqref{eqQM}. For $J_{ex}+J_{in} < J_H \approx 14.7$, the only attractor of the system is the fixed point and for $J_{ex}+J_{in} > J_H $ --- a limit cycle.  The white color in the figure corresponds to the regions where the above symmetric solutions are transverse stable. In the blue regions, these solutions are transverse unstable and thus here the system \eqref{eq_rv2} has no symmetric attractors. In the next section, we analyze symmetry-broken  solutions that appear in the blue regions of Fig.~\ref{fig_JexJinB}. 

\section{\label{sec4}Non-symmetric solutions and bifurcations}

The behavior of the interacting neural populations is different for inhibitory ($J_{ex}<0$) and excitatory ($J_{ex}>0$) couplings. Below we present the analysis of non-symmetric solutions for these two different cases in separate sections. 

\subsection{\label{sec4a} Inhibitory coupling}

\begin{figure*}
\centering\includegraphics{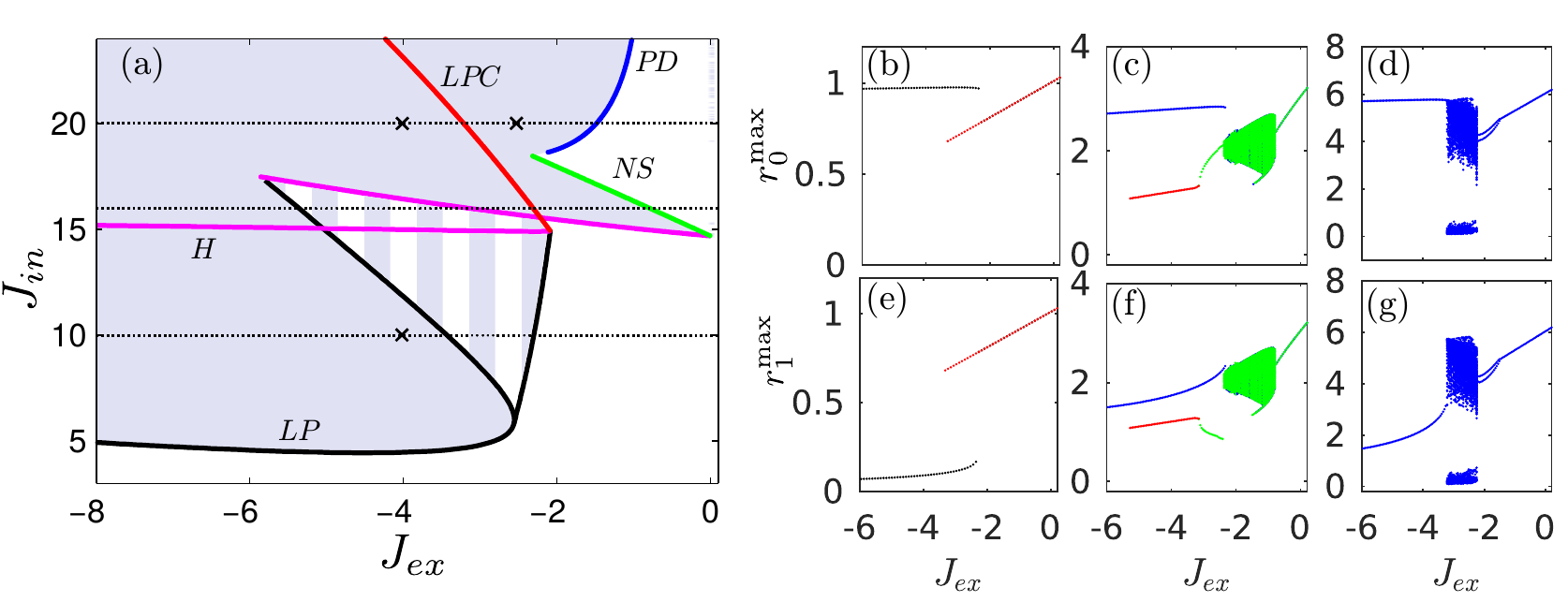}
\caption{\label{fig:JexJin} (a) Two-parameter $(J_{ex},J_{in})$ bifurcation diagram of the macroscopic model \eqref{eq_rv2} for inhibitory coupling between populations at fixed parameters $\bar{\eta} = 0$, $V_{th} = 50$ and $\Delta=1$. The white, colored and striped areas  define respectively the  monostable symmetric states, the non-symmetric attractors and the bistable states of the system. Continuous curves of different color, marked by acronyms, represent bifurcations: \textit{LP} -- limit point (black), \textit{H} -- Hopf (magenta), \textit{NS} -- Neimark-Sacker (green), \textit{LPC} -- limit point of cycles (red) and \textit{PD} -- period doubling (blue). The right-hand-side of the figure shows one-parameter bifurcation diagrams. They are constructed as a continuation of the solutions of Eqs.~\eqref{eq_rv2} via gradual change of the parameter $J_{ex}$ from zero to $-6$ at three different fixed values of $J_{in}$: 10 [(b) and (d)],  $16$ [(c) and (f)] and $20$ [(d) and(g)]. The latter values are presented in (a) by horizontal dotted lines. The crosses on these lines denote the values of the parameters at which the dynamics of the system are demonstrated in the subsequent figures in more details. The black points in (b) and (e) as well as red points in (b)-(f) show the stationary values of the spiking rates in different populations. The blue and the green points in (c)-(g) show the local maxima of the oscillating spiking rates.}
\end{figure*}

The bifurcation diagram of non-symmetric solutions in the region of parameters $(J_{ex},J_{in})$ relevant to the inhibitory coupling between populations  is shown in Fig.~\ref{fig:JexJin}(a). Here as well as in Fig.~\ref{fig_JexJinB} the white area corresponds to monostable symmetric states. In the colored region, there are non-symmetric attractors, while in the striped area the system exhibits coupling-induced bistability. Here, depending on the initial conditions, the system can approach either symmetric or non-symmetric state. 

The variation of the coupling strengths within and between the populations leads to a rich variety of bifurcations, including limit point (LP), Hopf (H), Neimark-Sacker (NS), limit point of cycles (LPC), and period doubling (PD) bifurcations. In Fig.~\ref{fig:JexJin}(a) these bifurcations are presented by colored curves marked with corresponding acronyms.    
A more detailed visualization of the bifurcations is given in the right-hand side of Fig.~\ref{fig:JexJin}. Here  the local maxima $r_0^{max}$ and $r_1^{max}$ of spiking rates of different populations are shown as functions of a smoothly varying coupling strength $J_{ex}$ for three different fixed values of $J_{in}=10, 16$ and $20$. These values are presented in the panel (a) by dotted horizontal lines.  

For low coupling strength $J_{in}=10$, the non-interacting populations ($J_{ex}=0$) have a symmetric fixed point attractor with equal time-independent spiking rates $r_0=r_1=\textit{const}$. They are marked in panels (b) and (e) by red points. The increase of the inhibitory coupling between the populations results in symmetry breaking through a limit point bifurcation.  The symmetric  fixed point attractor (red dots) exists in the interval $J_{ex} \in (-3.31, 0)$. The non-symmetric fixed point attractor (black dots) with different stationary spiking rates  in different populations, $r_0\ne r_1$, is in the  interval $J_{ex} \in (-6, -2.35)$. Following Ref.~\cite{olmi2010} we refer to this solution as \emph{a splay state}. The splay state is characterized  by the absence of any collective dynamics, since both the spiking rates and mean fields of populations are constant. In the interval $J_{ex} \in (-3.31, -2.35)$, the system exhibits bistability with coexisting symmetric and non-symmetric fixed point attractors.

More interesting symmetry breaking scenarios are observed at higher coupling strength $J_{in}>J_H\approx 14.7$, when the synchronization between neurons within the isolated populations ($J_{ex}=0$) causes macroscopic limit cycle oscillations of their spiking rates and mean fields. In  panels (c) and (f), we fix $J_{in}=16$ and continue the symmetric limit cycle solution to the region $J_{ex}<0$ (green dots). With the increase of the inhibitory coupling between the populations, the symmetry of the limit cycle is conserved up to the value $J_{ex} = -0.76$. At this point the system undergoes the Neimark-Sacker bifurcation. Further increase of the coupling strength $|J_{ex}|$  results in quasiperiodic oscillations, which then transform into periodic oscillations and finally become the symmetric fixed point attractor (red dots) following the Hopf bifurcation at $J_{ex} = -3.15$. This symmetric fixed point attractor disappears via limit point bifurcation at $J_{ex} = -5.28$. In addition to the above solutions, the system has a non-symmetric limit cycle attractor presented by blue dots. The attractor appears via a limit point cycle bifurcation at $J_{ex} = -2.41$ and exists for any $J_{ex} < -2.41$. This solution is most interesting since it represents \emph{a chimera-like state}. Here the spiking rate in one of the populations oscillates at high amplitude, while the spiking rate of other population has a low amplitude of oscillations. Below we discuss such solutions in more details. Note that the system is bistable in the interval $J_{ex} \in (-5.28, -2.41)$. In this interval, the chimera-like state may coexist either with a symmetric fixed point or with a non-symmetric limit cycle or with quasiperiodic oscillations.      

Further increase of the coupling strength $J_{in}$ within the populations revokes the bistability. 
In panels (d) and (g) this is demonstrated for the fixed value $J_{in} = 20$. The continuation of the symmetric limit cycle solution  into the region $J_{ex}<0$ shows that the system has no coexisting attractors for any $J_{ex}<0$. Unlike in the previous case, here the symmetry breaking appears through a period doubling bifurcation, which occurs at $J_{ex}=-1.46$. Latter this bifurcation leads to a chaotic regime in which high and low activities of neurons change irregularly between the populations. The microscopic dynamics of the system in this regime will be presented below. The chaotic regime is replaced by a periodic solution via a limit point cycle bifurcation at $J_{ex} =-3.25$. The non-symmetric periodic solution that exists for $J_{ex} <-3.25$ represents the above-mentioned chimera-like state. 

\begin{figure}
\centering\includegraphics{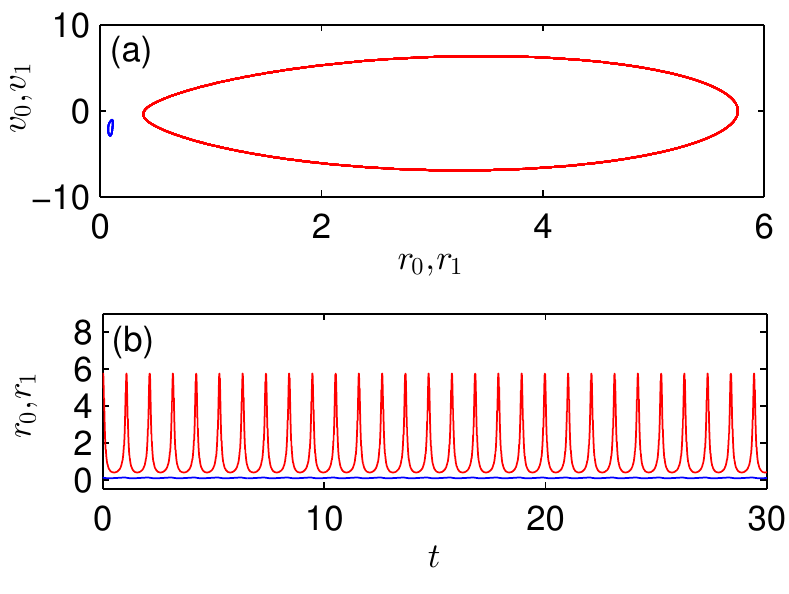}
\caption{\label{fig_chimera_macro} Chimera-like solution of the macroscopic model \eqref{eq_rv2}  for inhibitory coupling between populations at $(J_{ex},J_{in})=(-4,20)$, and other parameters the same as in Fig.~\ref{fig:JexJin}. (a) Projections of the solution to the plains $(r_0,v_0)$ and $(r_1,v_1)$ are presented by different colors. (b) Dynamics of the spiking rates in different populations also shown by different colors.} 
\end{figure}

To better visualize the chimera-like state, in Fig.~\ref{fig_chimera_macro} we fix $J_{ex}=-4$ and show the projections of the trajectory to the $(v_0,r_0)$ and $(v_1,r_1)$ plains as well as the dynamics of the spiking rates $r_0(t)$ and $r_1(t)$ in different populations. We see that the spiking rates in different populations oscillate with considerably different amplitudes. In one of the populations the spiking rate is close to zero (blue curve in the figure) and its oscillations are almost unremarkable on the scale of variation of the spiking rate of other population (red curve).  

\subsubsection{Simulation of the microscopic model}

Now we consider the problem of two interacting QIF neuron populations on the microscopic level.  Numerical simulations of the microscopic model Eqs.~\eqref{modelC} are interesting for two reasons. First, we can verify whether the macroscopic Eqs.~\eqref{eq_rv2} derived in the thermodynamic limit $N\to\infty$  predict well the dynamical regimes of a finite-size network. Second, such simulations allow us to observe the behavior of individual neurons for any particular macroscopic regime.    

Direct numerical integration of Eqs.~\eqref{modelC} is problematic because the membrane potential
of the QIF neuron tends to infinity at the moment when the neuron fires. This problem can be avoided by the change of variables
\begin{equation}
V_{j,k} = \tan(\theta_{j,k}/2)
\end{equation}
that transforms the QIF neurons into theta neurons, where $\theta_{j,k}$ is the phase of the $j$th neuron in the $k$th population. Then the model~\eqref{modelC} in the theta representation reads as follows:
\begin{equation}
\dot{\theta}_{j,k}=(1-\cos \theta_{j,k})+(1+\cos \theta_{j,k})(\eta_{j}+ \mathcal{I}_{k}). \label{eq_theta1}
\end{equation}
When the QIF neuron fires, its membrane potential approaches infinity, $V_{j,k} \to \infty$, and then its value is reset to minus infinity, $V_{j,k} \to -\infty$. In the theta representation,
this process is smooth: the phase $\theta_{j,k}$ simply crosses the value $\pi$. 

The parameters $\eta_{j}$, satisfying the Lorentzian distribution Eq.~\eqref{eq_lorz}, were generated deterministically by using formula $\eta_{j}=\bar{\eta}+\Delta \tan\left[(\pi/2)(2j-N-1)/(N+1)\right]$,  where $j=1,\ldots, N$ and $\Delta=1$. Such a numeration of neurons means that the isolated neurons with the index $j<j_c = (N+1)/2-(2N+1)\arctan(\bar{\eta})/\pi$ are excitable and the neurons with the index $j>j_c$ are spiking. At each step of integration of Eqs.~\eqref{eq_theta1}, the synaptic variables~\eqref{Sdef} were estimated as $S_k(t) = dN_k^S/N$, where $dN_k^S$ is the number of neurons in the $k$th population whose phases are in the interval $\theta_{j,k} \in [2 \arctan(V_{th}),\pi]$. Similarly, the firing rates were estimated as $r_{k}=dN_k^r/(Ndt)$, where $dN^r_k$ is the number of neurons in the $k$th population whose phases are in the interval $\theta_{j,k} \in (\pi-2 d t, \pi)$. Such estimations are based on the assumption that the time step $dt$ is small and thus the phase speed of neurons close to the firing phase $\theta=\pi$ can be approximated as $\dot{\theta}_{j,k} \approx 2$. Because of the finite number of neurons, the quantities $dN_k^r$ fluctuate in time and the firing rates vary nonsmoothly. For better visualization, we smoothed these quantities by using a moving average with a time window of the size $\delta t=5\cdot10^{-2}$.
\begin{figure*}
\centering\includegraphics[width=2.05\columnwidth]{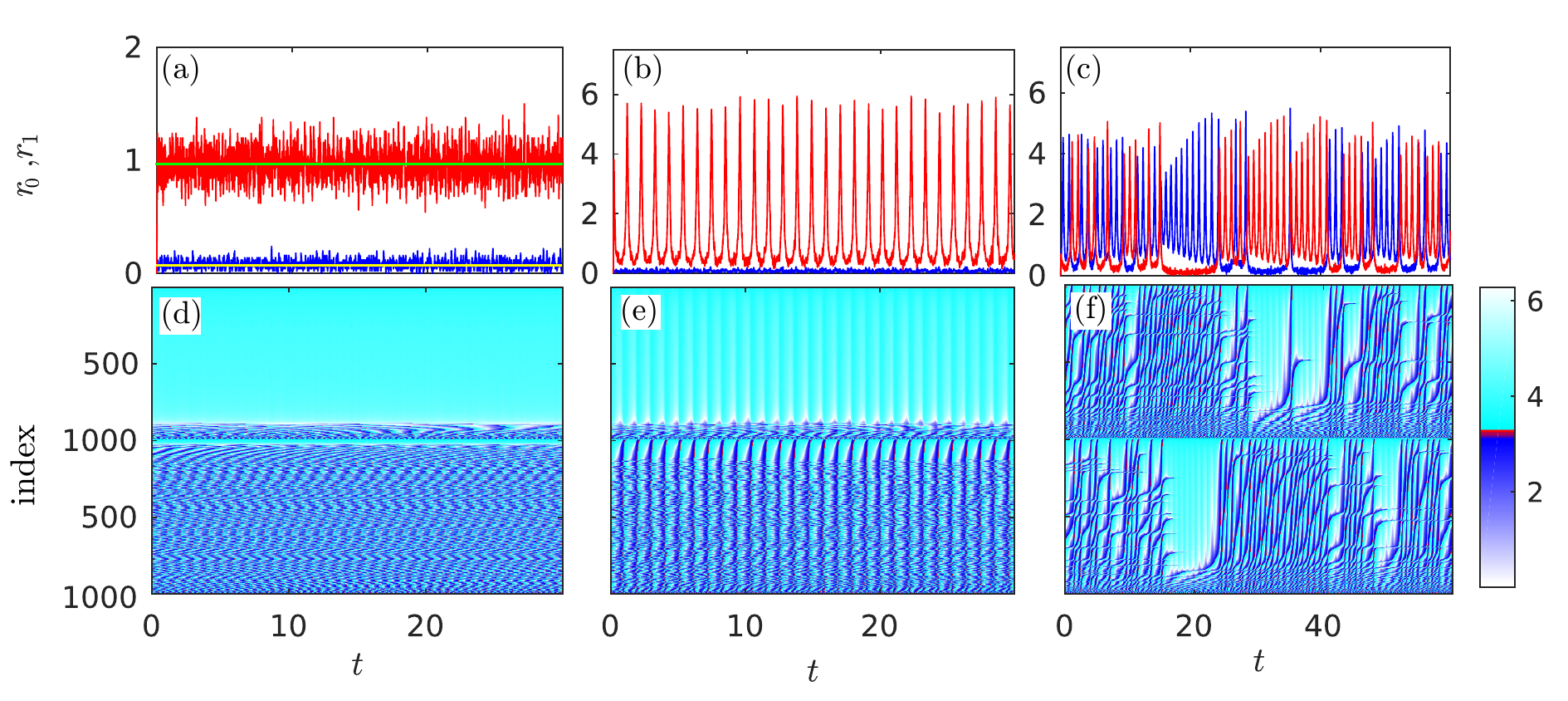}
\caption{\label{fig:phase} Modeling of two inhibitory coupled populations of QIF neurons,  according to the microscopic Eqs.~\eqref{eq_theta1}. The number of neurons in each population is  $N=1000$.  (a)-(c) Dynamics of spiking rates of different populations shown by different colors. (d)-(f) Microscopic dynamics of the phases of neurons in both populations.  Three columns of the figure correspond to three different sets of the coupling strengths $(J_{ex},J_{in})$: $(-4,10)$  [(a) and (d)], $(-4,20)$  [(b) and (f)] and $(-2.5,20)$  [(c) and (f)]. Other parameters are the same  as in Fig.~\ref{fig:JexJin}. Two horizontal lines in (a) show the solution of the macroscopic model \eqref{eq_rv2}.}
\end{figure*}

The three columns in Fig.~\ref{fig:phase} show the dynamics of the  microscopic model for three different dynamical regimes defined by  different choices of the coupling strengths $(J_{ex}, J_{in})$. In order from left to right, they are $(-4,10), (-4,20)$ and $(-2.5, 20)$. These values were marked in Fig.~\ref{fig:JexJin}(a) by crosses. The upper (a)-(c) and bottom (d)-(f) panels in Fig.~\ref{fig:phase} show, respectively the dynamic of the spiking rates and  phases of individual neurons in both populations.

The panels (a) and (d) correspond to the splay state. According to the macroscopic model, this state is characterized by  stationary spiking rates, which are different in different populations. For the given values of the parameters, the macroscopic model predicts the values of the spiking rates $r_0=0.09$ and $r_1=0.98$. In panel (a) they are shown by horizontal lines. The spiking rates obtained from the microscopic model fluctuates around these predicted values. The fluctuations occur due to the finite size of the network. In the population with a small spiking rate, almost all neurons are quenched, while in the population with a larger spiking rate almost all neurons spike, however, their spikes are incoherent and they do not produce any macroscopic oscillations. Note that for the given value $\bar{\eta}=0$, the half of the isolated ($J_{ex}= J_{in}=0$) neurons in each population are spiking and another half are excitable.   

The chimera-like state reproduced by the microscopic model is shown in panels (b) and (e). The values of the parameters are the same as in Fig.~\ref{fig_chimera_macro}, where this state is demonstrated via the macroscopic model. Comparing Figs.~\ref{fig_chimera_macro}(b) and \ref{fig:phase}(b), we see that the dynamics of the spiking rates derived from the microscopic and macroscopic models are in good agreement. In the chimera-like state, the majority of neurons in one of the populations are quenched and their spiking rate is close to zero, while in another population  the majority of neurons spike synchronously and produce large amplitude oscillations of the spiking rate. 

Finally, in panels (c) and (f)  we demonstrate the solutions of microscopic equations for a chaotic chimera-like state. Here synchronous spiking erratically jumps from one to another population. For the given values of the parameters, the macroscopic model exhibits similar behavior (not shown).

\subsection{\label{sec4b} Excitatory coupling}

We turn now to the situation when couplings within and between the populations are both excitatory, $J_{in}>0$ and $J_{ex}>0$. In this case the bifurcation scenarios are less diverse than for the inhibitory coupling considered above. Now the qualitative change of solutions is mainly defined by only two bifurcations, namely, the Neimark-Sacker and branch point of cycles (BPC) bifurcations.

\begin{figure}
\centering\includegraphics{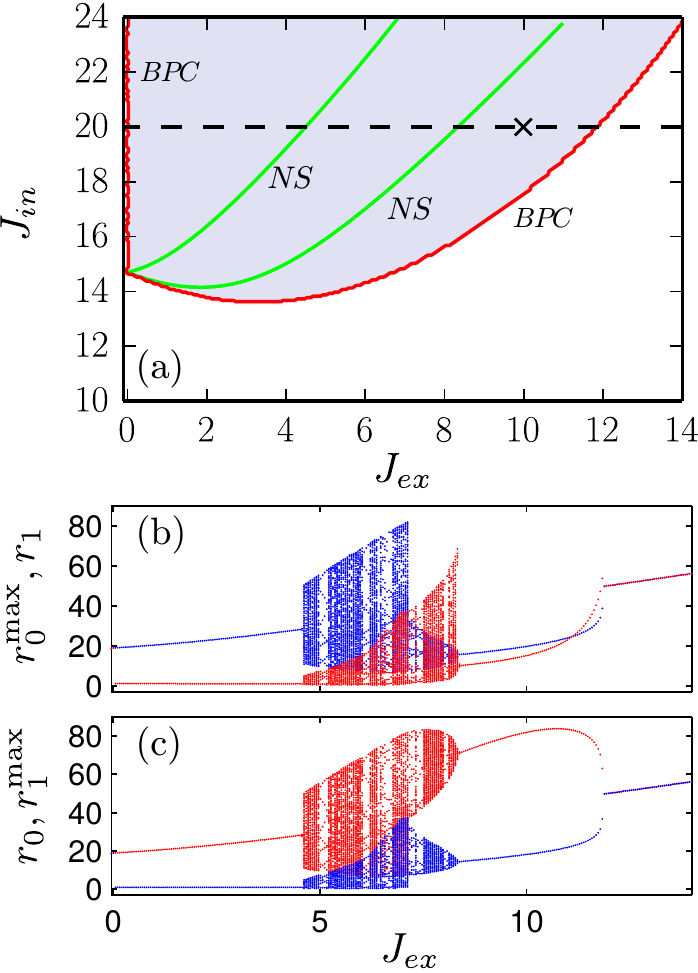}
\caption{\label{fig_Jex_Jin_pos} (a) Two-parameter $(J_{ex},J_{in})$ bifurcation diagram of the macroscopic model \eqref{eq_rv2} for excitatory coupling between populations at fixed parameters $\bar{\eta} = 0$, $V_{th} = 50$ and $\Delta=1$. The white and colored areas  correspond respectively to the  monostable symmetric states and the non-symmetric attractors.   Continuous curves marked by acronyms, represent bifurcations: \textit{NS} -- Neimark-Sacker (green) and \textit{BPC} -- branch point of cycles (red). (b) and (c) One-parameter bifurcation diagrams constructed as a continuation of the solutions of Eqs.~\eqref{eq_rv2} via a gradual increase of the parameter $J_{ex}$ from zero to $14$ at fixed $J_{in}=20$. Blue and red points in (b) show, respectively the local maxima  of $r_0$ denoted as $r_0^{max}$ and corresponding values of $r_1$ when these maxima are attained. In (c) the red color corresponds to $r_1^{max}$,  while the blue color defines $r_0$.} 
\end{figure}

In Fig.~\ref{fig_Jex_Jin_pos}(a), we present the bifurcation diagram of the macroscopic model \eqref{eq_rv2} in the region of parameters $(J_{ex},J_{in})$ relevant to the excitatory coupling between populations. Here as well as in Fig.~\ref{fig:JexJin} the white area corresponds to monostable symmetric states and the colored region represents non-symmetric attractors. At the border of the colored region (red curve) the system  undergoes BPC bifurcation. At this bifurcation the symmetric limit cycle looses stability and there appears a symmetrical pair of stable non-symmetric limit cycles. The two green curves in the colored region indicate  Neimark-Sacker bifurcation. 

In panels (b) and (c), we fix $J_{in}=20$ and continue the solutions of Eqs.~ \eqref{eq_rv2} into the region $J_{ex}>0$. Unlike to Figs.~\ref{fig:JexJin}(b)-(g), here we plot not only the local maxima $r_0^{max}$ and $r_1^{max}$ of the spiking rates of different populations, but also the values $r_1$ and $r_0$, respectively, at the moments when the neighbor population has reached its local maximum. This representation allows us to distinguish antiphase periodic oscillations $r_0(t)=r_1(t+T/2)$, where $T$ is the period of the limit cycle. Such oscillations are common for the excitatory coupling. For fixed $J_{in}=20$, the antiphase periodic oscillations take place in the region $0<J_{ex}<4.5$. In the region $4.5<J_{ex}<8.3$, between two Neimark-Sacker bifurcations, the system exhibits complex dynamics, including high-periodic cycles, quasiperiodicity and chaos. Then for $8.3<J_{ex}<11.9$, the chimera-like state appears. This state is destroyed via a branch point of cycles bifurcation at $J_{ex}=11.9$, and for $J_{ex}>11.9$ the symmetric limit cycle oscillations are established.
\begin{figure}
\centering\includegraphics{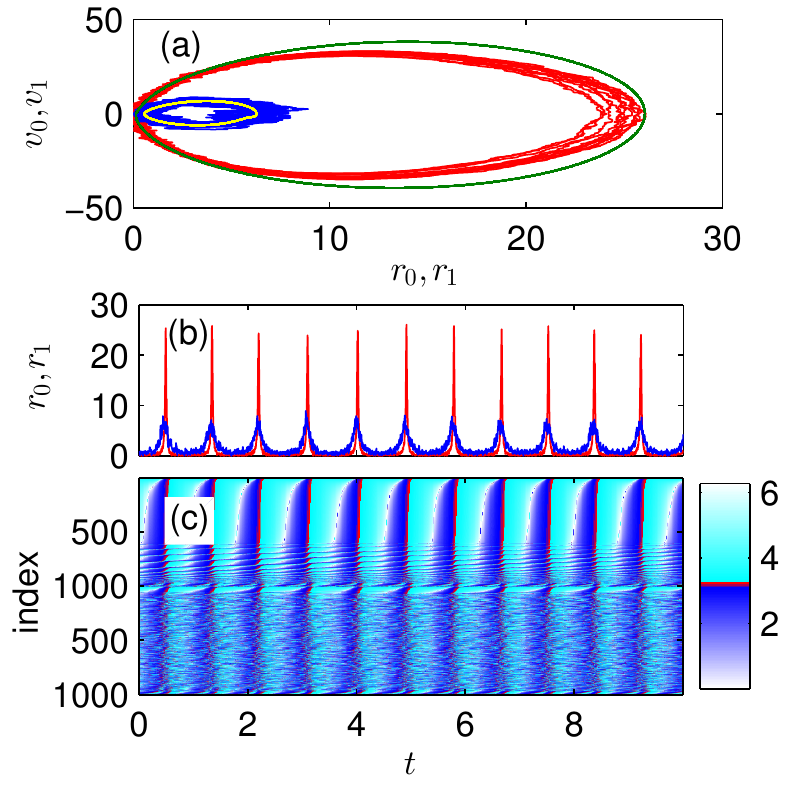}
\caption{\label{fig:r_v} Chimera-like state for excitatory coupling between populations. The microscopic Eqs.~\eqref{eq_theta1} were solved at $(J_{ex},J_{in})=(10,20)$, $N=1000$ and other parameters the same as in Fig.~\ref{fig_Jex_Jin_pos}.
(a) Projections of the solution to the plains $(r_0,v_0)$ and $(r_1,v_1)$. The smooth green and yellow closed curves show the solutions of the macroscopic model \eqref{eq_rv2}. (b) Dynamics of spiking rates in different populations. (c) Microscopic dynamics of the phases of neurons in both populations.}
\end{figure}

The microscopic dynamics of the system for the periodic chimera-like state at $(J_{ex},J_{in})=(10,20)$ [in Fig.~\ref{fig_Jex_Jin_pos}(a) these values are marked by a cross] is demonstrated in Fig.~\ref{fig:r_v}. In panel (a) the projections of the microscopic dynamics to the plains $(r_0,v_0)$ and $(r_1,v_1)$ are compared with the corresponding projections obtained from the macroscopic model. Again, we see that the macroscopic model predicts well the dynamics of the finite-size network. Panels (b) and (c) show the time-dependence of the spiking rates and phases of neurons, respectively, for both populations. Comparing Figs.~\ref{fig:r_v}(c) and \ref{fig:phase}(e), we reveal that the microscopic dynamics of the system in chimera-like state are different for the inhibitory and excitatory coupling between populations. For the inhibitory coupling, almost all neurons in one population spike in synchrony and in another population the majority of neurons are quenched. For the excitatory coupling, in both populations almost all neurons spike, but in one population, they spike synchronously and in another -- asynchronously. These two different microscopic dynamics lead to the similar macroscopic result: for one population,  the spiking rate exhibits large amplitude oscillations, while for another population it is close to zero.

\section{\label{sec5}Mean field coupling}

In this section, we additionally consider  another type of coupling between the populations. As well as above, we assume that the interaction within the populations is defined by the synaptic current \eqref{syn_curr1}, however the coupling between the populations is provided by mean field of membrane potential. Experimentally, such a situation can be imagined as a control problem. Assume we have two non-interacting populations of neurons and can separately measure their mean membrane potentials. Then we stimulate the first population by a signal proportional to the mean field measured from the second population and vice versa. In that case the total current $\mathcal{I}_{k}$ that defines all interactions between and within the populations in Eq.~\eqref{modelC} takes the form:
\begin{equation}
\mathcal{I}_{k} =  J_{in}  S_{k} V_{th}+J_{ex} v_{1-k},\label{Isyn1}
\end{equation}
where 
\begin{equation}
v_{k} =  \frac{1}{N}\sum_{l=1}^{N} V_{l,k} \label{vmean}
\end{equation}
is the mean membrane potential of the $k$th population. The macroscopic dynamics of this system are again described by Eqs.~\eqref{eq_rv2}, but the total current  is now defined by Eq.~\eqref{Isyn1}.

\begin{figure}
\centering\includegraphics{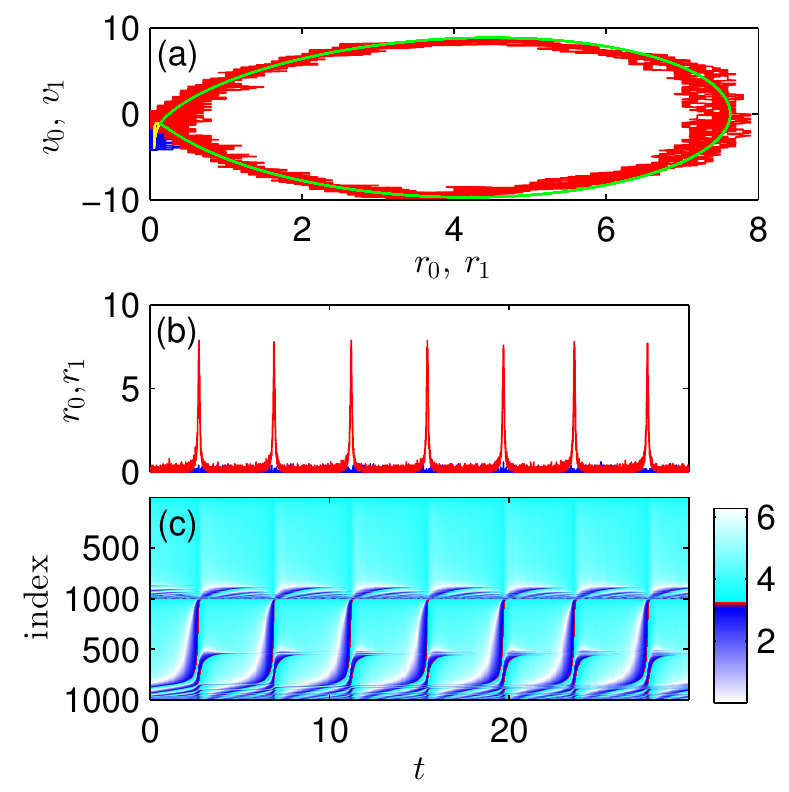}
\caption{\label{fig:mean_coupl} The same as in Fig.~\ref{fig:r_v}, but the coupling between populations is provided by mean field, according to Eqs.~\eqref{Isyn1} and \eqref{vmean}. The values of the parameters are: $J_{ex}=-2$, $J_{in}=23$,  $\bar{\eta} = -7$, $V_{th} = 50$ and $\Delta=1$. Dynamics of the spiking rates are smoothed by moving average method with a time window of the size $\delta t = 10^{-2}$.}
\end{figure} 
An example of chimera-like state obtained with an inhibitory mean field coupling between the populations is demonstrated in Fig.~\ref{fig:mean_coupl}. In numerical simulations of the microscopic model, the mean membrane potential~\eqref{vmean} were estimated by ignoring the contribution from neurons with extremely large values of membrane potential, when $V_{l,k}>200$. This allowed us to avoid the divergence of the sum~\eqref{vmean} due to the finite-size effect.  The dynamics of the system presented in Fig.~\ref{fig:mean_coupl}(b)-(c) is similar to that shown in Fig.~\ref{fig:phase}(b)-(e), which has been obtained with the inhibitory synaptic coupling.

\section{\label{sec6}Discussion}

In this paper, we analyzed the dynamics of a neural network consisting of two identical populations of quadratic integrate-and-fire neurons. Both populations are heterogeneous; they include inherently spiking and excitable neurons. The main part of the paper is devoted to the  synaptic coupling within and between the populations. The coupling is global and takes into account the finite width of synaptic pulses. We assumed that the interactions within the populations are excitatory,  and considered two cases, with the inhibitory and excitatory coupling between the populations. 

Using a recently developed reduction technique \cite{montbrio15} based on the Lorentzian ansatz, we  derived a macroscopic model that describes the neuron  spiking rates and mean membrane potentials in different populations. The model is defined by only four ordinary differential equations, which are exact in the limit of the infinite-size network. Such a simplification allowed us to perform a thorough bifurcation analysis of the system. In the parameter plane defined by the coupling strengths within and between the populations, we identified the areas where the symmetric solutions (with the identical dynamics in both populations) lose their stability and non-symmetrical solutions are established. Our analysis showed that the competition of neural interactions within and between the populations may lead to a rich variety of non-symmetric patterns, including a splay state, antiphase periodic oscillations, chimera-like states, also chaotic oscillations as well as bistabilities between various states. The most interesting non-symmetric pattern is the chimera-like state. Here the neurons in one population behave synchronously and produce high amplitude oscillations of the spiking rate, while the neurons in another population are quenched or desynchronized and their spiking rate is close to zero. The chimera-like state exists for both the inhibitory and excitatory synaptic coupling between the populations. In addition, we showed that the chimera-like state may appear when the synaptic coupling between the populations is replaced with a mean field coupling.

To verify the validity of the macroscopic model we performed numerical simulations of the microscopic model equations for a finite-size network. As a result, we were convinced that the macroscopic model predicts well the behavior of a finite-size network consisting of only thousand neurons in each population. Thus the macroscopic models of the type considered here are natural candidates for use in future large-scale brain simulations. Such models can be considered as an alternative to neural mass models \cite{Destexhe2009}, which are especially useful for understanding brain rhythms. The neural mass models also employ only several differential equations to describe the coarse-grained activity of large-scale neural networks, however, they are phenomenological in nature. In contrast, the approach discussed in this paper provides an exact macroscopic description of an underlying microscopic spiking neurodynamics.

\bibliography{symm_breake}

\end{document}